\documentclass{csmagclass2}														

\begin{document}		
\headings																					

\title{{ Entanglement-entropy study of phase transitions\\in six-state clock model }}
\author[1]{{R. KR\v CM\'AR\thanks{Corresponding author: roman.krcmar@savba.sk}}}
\author[1]{{A. GENDIAR}}
\author[2]{{T. NISHINO}}
\affil[1]{{Institute of Physics, Slovak Academy of Sciences, D\'ubravsk\'a c. 9, 845 11, Bratislava, Slovakia}}
\affil[2]{{Department of Physics, Graduate school of science, Kobe University, Kobe 657-8501, Japan}}

\maketitle

\begin{Abs}
The Berezinskii-Kosterlitz-Thouless (BKT) transitions of the six-state clock model on the square lattice 
are investigated by means of the corner-transfer matrix renormalization group method. A classical analog 
of the entanglement entropy $S( L, T )$ is calculated for $L \times L$ square system up to $L = 129$, as 
a function of temperature $T$. The entropy exhibits a peak at $T = T^*_{~}( L )$, where the temperature 
depends on both $L$ and the boundary conditions. Applying the finite-size scaling to $T^*_{~}( L )$ and 
assuming presence of the BKT transitions, the two distinct phase-transition temperatures are estimated 
to be $T_1^{~} = 0.70$ and $T_2^{~} = 0.88$. The results are in agreement with earlier studies. It should 
be noted that no thermodynamic functions have been used in this study. 
\end{Abs}


\section{Introduction}

The classical XY model on uniform planar lattices does not exhibit the `standard' type of order when the 
temperature $T$ is finite since the system possesses the continuous $O( 2 )$ symmetry~\cite{mermin}.
The special type of the order that does not break the symmetry can, however, exist at finite temperature 
and is known as the topological order~\cite{berezinskii, kosterlitz}. The phase transition between the 
topological phase and the high-temperature paramagnetic (or disordered) phase is the so-called BKT phase transition.

Introduction of anisotropy or discreteness is relevant to the thermodynamic properties of the system. 
The $q$-state clock model is one of the well-known examples, where on each lattice point there 
is a vector spin pointing to $q$ different directions, which differ by the angle $2\pi / q$. Since there is
no continuous symmetry, existence of the standard ferromagnetic order is allowed at low, but finite,
temperature. An early renormalization-group (RG) study on such a system by Jos\'e and Kadanoff suggested 
existence of a critical area with a finite temperature width~\cite{jose}, which is separated from ordered 
and disordered phases by the BKT phase transition~\cite{berezinskii,kosterlitz}. It has been known that 
such a phase structure exists for ferromagnetic $q$-state clock models when $q \ge 5$. It is known that 
within that temperature region $T_1^{~} < T < T_2^{~}$ the correlation function shows a power-law decay, 
and the system is critical. 

In this article we consider the ferromagnetic six-state ($q = 6$) clock model on the square 
lattice, as a representative case where the BKT transition can be observed. The Hamiltonian 
of the system, 
$H = - \sum_{i, j}^{~} \left[
		\cos\left( \theta_{i,j}^{~} - \theta_{i+1,j}^{~} \right)
	      + \cos\left( \theta_{i,j}^{~} - \theta_{i,j+1}^{~} \right)
	                       \right] $, 
where $\theta_{i,j}^{~} = \frac{2 \pi k}{q}$ denotes a discrete angle variable for $k = 0, 1, 2, \dots, q - 1$ on 
the lattice site with coordinates $i$ and $j$ 
in the $L \times L$ square lattice. When the temperature $T$ is high enough, the thermal equilibrium state 
is disordered, and each direction is chosen with equal probability. When $T$ is low enough, the state is ordered, 
i.e., one of the six directions is spontaneously chosen in the thermodynamic limit. 

Entanglement entropy, which quantifies the bipartite quantum entanglement, is one of the 
fundamental values in information physics, and has been used for analyses of one-dimensional (1D) 
quantum systems~\cite{osborne, vidal, franchini}.
Through the quantum-classical correspondence formulated by means of discrete path-integral in
imaginary time, such as the Trotter-Suzuki decomposition~\cite{trotter, suzuki}, it is also possible 
to introduce a classical analog of the entanglement entropy for two-dimensional (2D) classical 
lattice systems~\cite{tagliacozzo,krcmar4}. A profit of using this classical analog is that it enables to
detect thermal phase transitions directly, without considering the type of the order parameter or without
taking derivatives of thermodynamic functions, including the free energy~\cite{krcmar1, krcmar2, krcmar3}. 
Universality of the phase transition 
can also be identified by estimating the central charge through the finite-entanglement 
scaling~\cite{tagliacozzo,krcmar1, krcmar3}.

In this article, the entanglement-entropy analysis is used for the first time in attempt to identify
the BKT transition. We calculate the entanglement entropy $S( L, T )$ of the 
six-state clock model on square-shaped systems of the linear sizes up to $L = 129$, and investigate the
phase  transition by means of temperature dependence in $S( L, T )$. For this purpose, we employ the 
Corner Transfer Matrix Renormalization Group (CTMRG) method~\cite{tomotoshi}.

\section{Numerical Results}

\begin{figure}
\includegraphics[width=0.45\textwidth,clip]{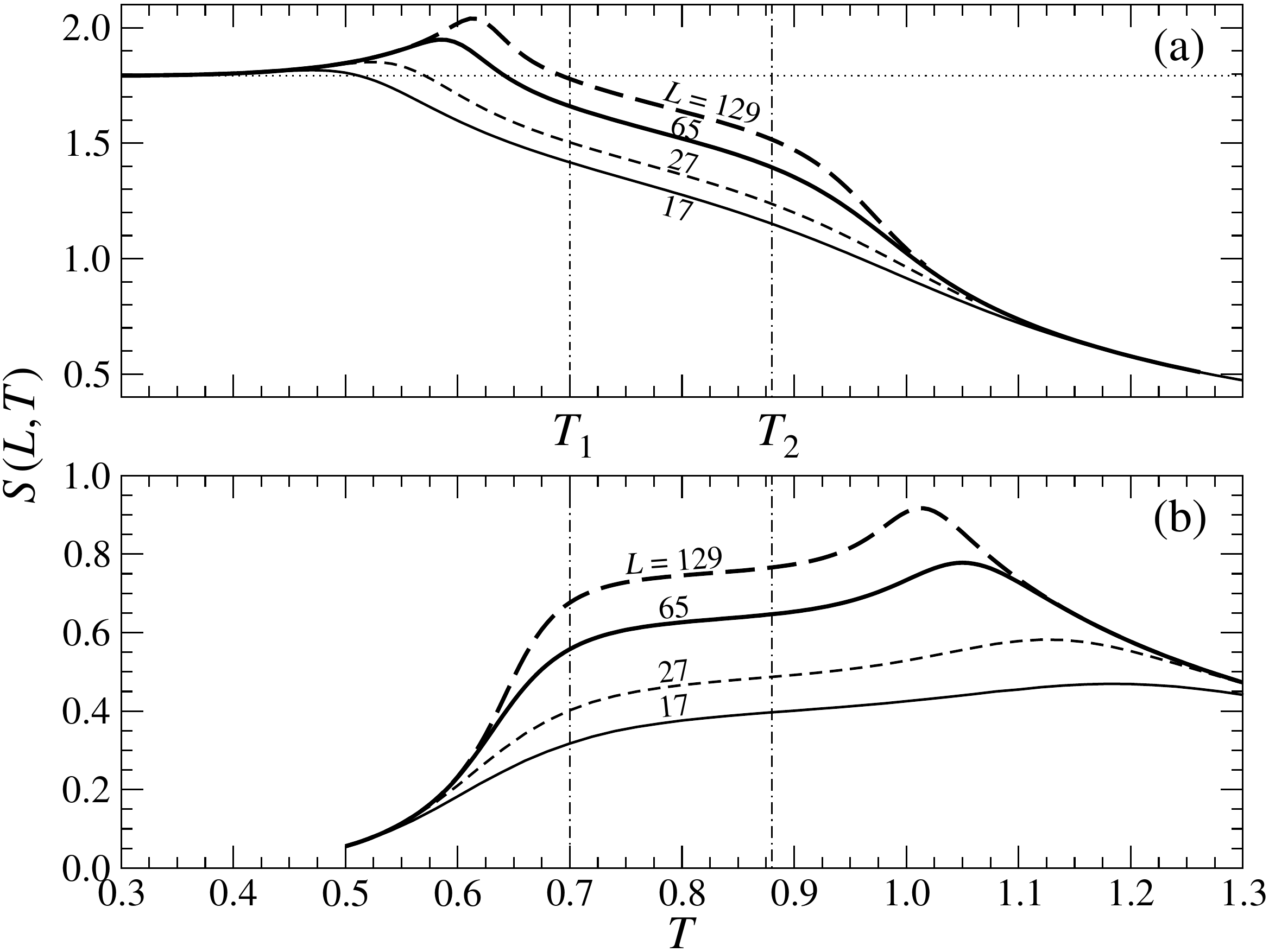}
\hfill
\includegraphics[width=0.45\textwidth,clip]{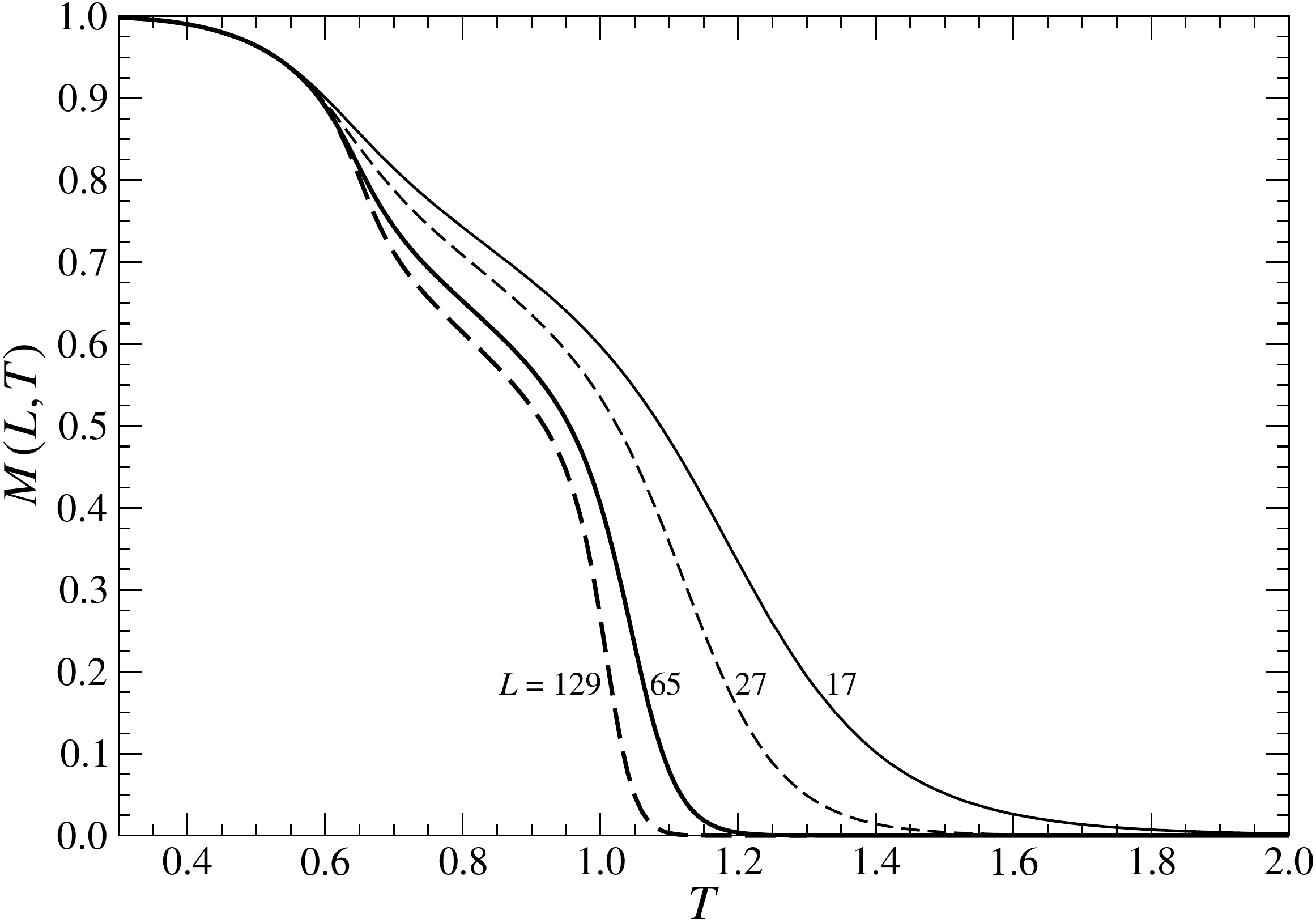}
\caption{Left: 
Temperature dependence of magnetization $M( L, T ) = \langle \cos ( \theta_{\rm c}^{~} ) \rangle$ 
measured at the center of the square shape system for $L = 17, 27, 65,$ and $129$ with fixed boundary conditions.
Right: Entanglement entropy $S( L, T )$ calculated with
(a) the free boundary conditions and (b) the fixed boundary conditions. The horizontal dotted line in (a) is 
the asymptotic value of $S( L, 0 ) = \ln \, 6$. The two vertical dot-dashed lines at $T_1^{~} = 0.70$ and 
$T_2^{~} = 0.88$ show the estimated transition temperatures by means of finite-size scaling, as discussed bellow.
The sizes of the square lattice are specified by the type of the lines; $L = 17$ (full thin line),  
$L = 27$ (dashed thin line), $L = 65$ (full thick line), and $L = 129$ (dashed thick line).}
\label{fig:entropy}
\end{figure}

Left panel of Fig.~\ref{fig:entropy} shows the temperature dependence of the magnetization 
$M( L, T ) = \langle \cos ( \theta_{\rm c}^{~} ) \rangle$
at the center of the square-shaped system, 
here denoted by the suffix ${\rm c}$. System is under the fixed boundary conditions. 
We have chosen the lattice sizes $L = 17, 27, 65$, and $129$. Having analyzed the magnetization profiles, 
it is non-trivial to determine the BKT transitions accurately. We, therefore, focus our attention on the  
form of the entanglement entropy $S( L, T )$ in order to detect two 
effective temperatures $T_1^{*}( L )$ and $T_2^{*}( L )$ for 
each lattice size $L$, which can be further used to analyze the BKT transitions.

The temperature dependence of the $S( L, T )$ is shown on the right panel of Fig.~\ref{fig:entropy} for
various sizes of the square lattice 
with $L = 17, 27, 65$, and $129$. We consider two different boundary conditions.
The upper panel shows $S( L, T )$ for the free boundary conditions. The entropy has a peak (maximum) at
$T_1^{*}( L )$, which increases with $L$. Above $T_1^{*}( L )$ the entropy decreases with $T$,
where there is a shoulder on the higher-temperature side, which carry an unclear signatures of $T_1^{*}( L )$.
(Notice that $S( L, T = 0 ) = \ln \, 6$.) If the fixed boundary conditions are imposed, shown in the lower panel
(b), there is a shoulder for each $L$ in the lower-temperature side and a new peak at $T_2^{*}( L )$, which
is a decreasing function of $L$. (Notice that $S( L, 0 ) = 0$.)

The BKT transition temperatures $T_1^{~}$ and $T_2^{~}$ can be obtained by applying the finite-size scaling
for $T_1^{*}( L, T )$ and $T_2^{*}( L, T )$, respectively, toward the thermodynamic limit $L \rightarrow \infty$.
Now, we check this conjecture. It has been accepted that the correlation length $\xi$ around the BKT 
phase-transition temperature $T_{\rm C}^{~}$ is asymptotically ($L \to \infty$) expressed to 
be $\xi \, \propto \, \exp\left( const. \frac{\sqrt{T_{\rm C}^{~}}}{\sqrt{ \left| T - T_{\rm C}^{~} \right|}} \right) \,$ . 
If the system size $L$ is smaller than 
$\exp\left( const. \frac{\sqrt{T_{\rm C}^{~}}}{\sqrt{ \left| T - T_{\rm C}^{~} \right|}} \right)$, the correlation 
length $\xi$ is effectively suppressed down to $L$. Under such a geometrical constraint, 
it is possible to introduce an effective temperature $T^{*}_{~}( L )$ that satisfies 
$L \, \propto \, \exp\left( const. \frac{\sqrt{T_{\rm C}^{~}}}{\sqrt{ \left| T^{*}_{~}( L ) - T_{\rm C}^{~} \right| }} \right) \,$ .
Solving this relation with respect to $T^{*}_{~}( L )$, we obtain 
$T^{*}_{~}( L ) = T_{\rm C}^{~} + \frac{\alpha}{{\left[ \ln ( \beta L ) \right]}^2_{~}} \,$ ,
where $\alpha$ and $\beta$ are appropriate constants. Since the entanglement entropy is 
almost proportional to the logarithm of the correlation length, an analogous consideration is applied
to the entanglement entropy in the following.

\begin{figure}
\includegraphics[width=0.45\textwidth,clip]{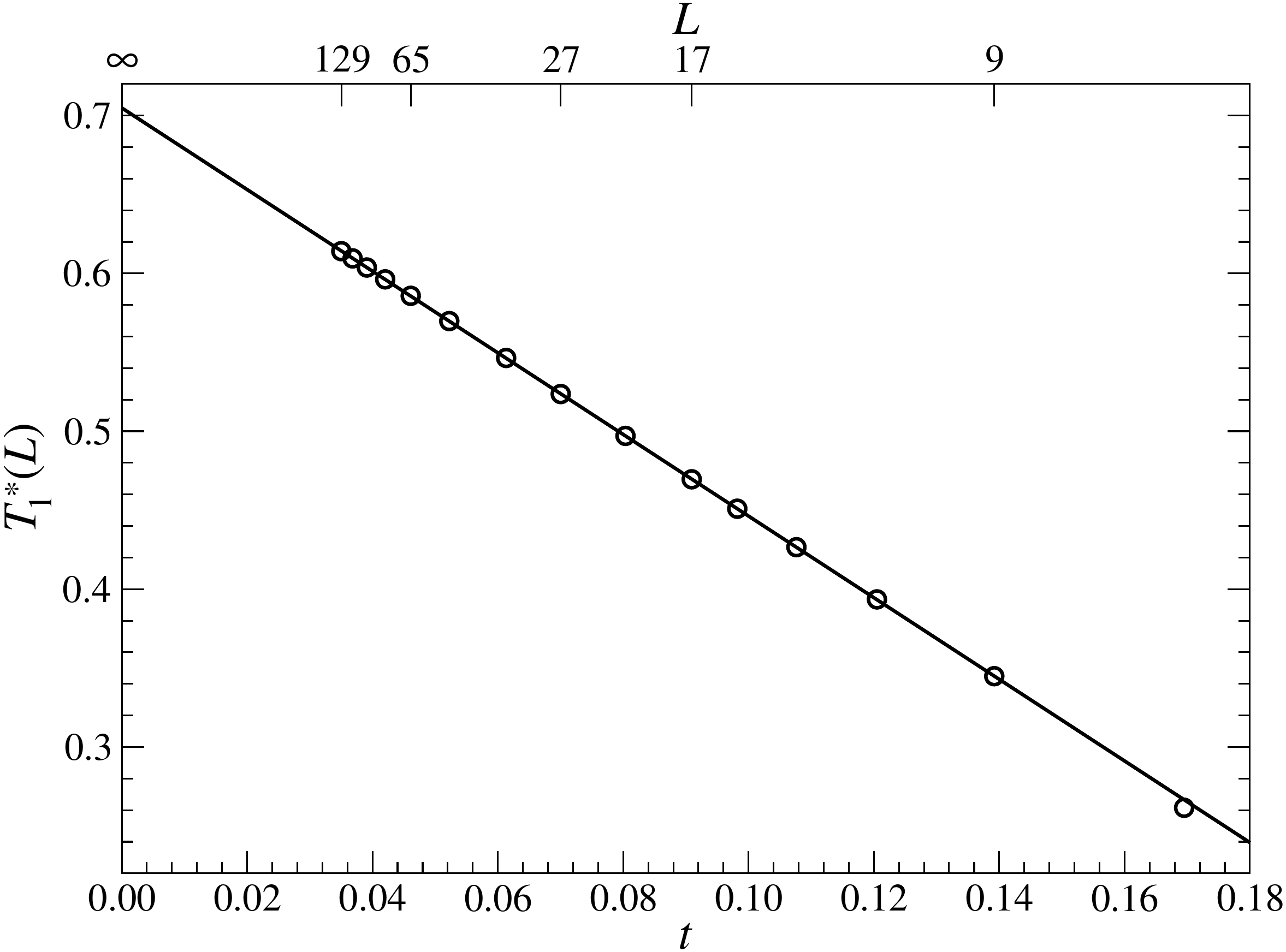}
\hfill
\includegraphics[width=0.45\textwidth,clip]{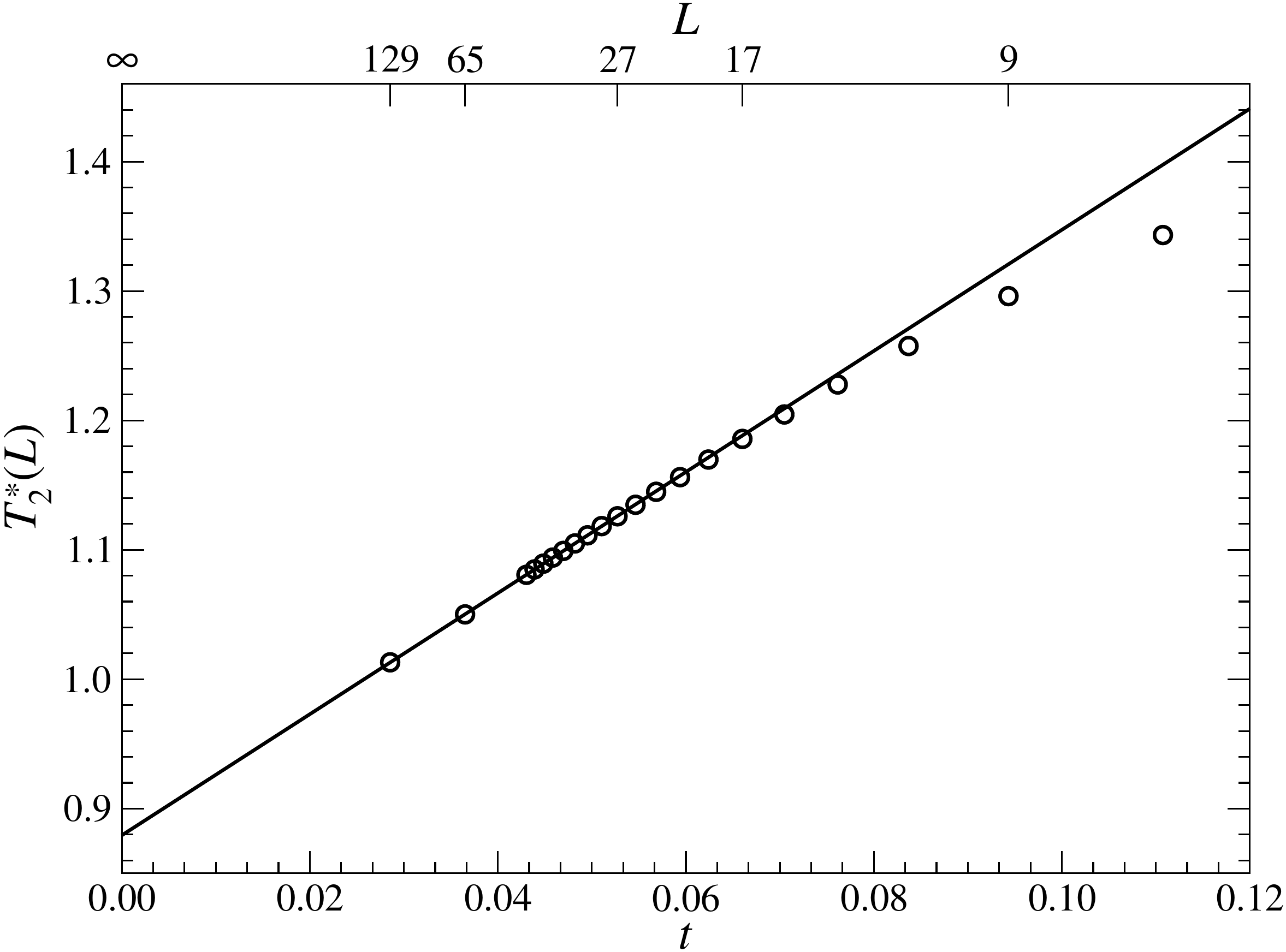}
\caption{Left: Finite-size scaling for the peak position $T_1^{*}( L )$ with respect to $t$, 
which draws $T_1^{*}( \infty ) = 0.70$. Right: Finite-size scaling for 
$T_2^{*}( L )$ with respect to $t$, which draws $T_2^{~} = 0.88$.}
\label{T1}
\end{figure}

Figure~\ref{T1} shows $T_1^{*}( L )$ and $T_2^{*}( L )$ with respect to
$t = \frac{1}{{\left[ \ln (\beta L) \right]}^2_{~}} \,$ (as well as with respect to $L$ in the logarithmic scale).
On the left graph, $T_1^{*}( L )$ is plotted for $\beta = 1.62$, which was determined to reach linearity
for low values of $t$ and the slope corresponds to $\alpha = - 2.58$. Under this parameterization of
$\alpha$ and $\beta$, we estimated the lower-temperature BKT transition to be $T_1^{*}( \infty ) = 0.70$. 
The right graph shows the analogous analysis for $T_2^{*}$ with respect to $t$, where the parameters
were found to be $\alpha = 4.68$ and $\beta = 2.88$. Finally, the higher-temperature BKT transition
is determined by the extrapolation to be $T_2^{~}( \infty ) = 0.88$.

\section{Conclusions}

We have studied the six-state clock model by means of the CTMRG method and observed the
temperature dependence of the entanglement entropy $S( L, T )$ with $L$ being the size of
the square lattice. If imposing the free boundary conditions, the entanglement entropy
exhibits the peak at $T_1^{*}( L )$, which is an increasing function of $L$. On the other
hand, the fixed boundary conditions results in the other peak of the entanglement entropy at
$T_2^{*}( L )$, which is the decreasing function of $L$. According to the BKT form of the
finite-size correction, the scalings applied to $T_1^{*}( L )$ and $T_2^{*}( L )$
draw the final results of the lower-temperature BKT transition $T_1^{*}( \infty ) = 0.70$ and
the higher-temperature BKT transition $T_2^{*}( \infty ) = 0.88$. These values agree with the
transition temperatures reported so far~\cite{lapilli, chatelain, kumano}. 
For comparison, the most recent Monte Carlo result by
Kumano {\it et al.} gives $T_1^{~} = 0.700(4)$ and $T_2^{~} = 0.904(5)$, 
which is based on the response to twist boundary conditions up to the size $L = 256$~\cite{kumano}.

\section{Acknowledgments}
This work was supported by the projects APVV-16-0186 (EXSES) and VEGA-2/0123/19.
T.~N. and A.~G. acknowledge the support of Grant-in-Aid for Scientific Research.
R.~K. acknowledges the support of Japan Society for Promotion of Science P12815.

\end{document}